\documentclass[aps,prl,twocolumn,showpacs,nobibnotes,reprint]{revtex4-1}
\usepackage{graphics,graphicx,amsfonts,amsmath,amsbsy,amssymb,color}
\usepackage{verbatim}  
\usepackage{psfrag}  
\usepackage{tabularx}
\usepackage{xmpmulti}
\usepackage{hyperref}
\usepackage{marginnote}
\usepackage{subfigure}
\usepackage{paracol}
\usepackage{xparse}
\usepackage{layouts}
\usepackage{afterpage}
\usepackage{braket}
\usepackage{paralist}
\usepackage{bm}
\usepackage{url}
\usepackage[normalem]{ulem}
\usepackage{color}
\usepackage[colorinlistoftodos]{todonotes}
\usepackage[american]{babel} 

\newcommand{\refeq}[1]{{Eq.~(\ref{#1})}}
\newcommand{\reffig}[1]{{Fig.~\ref{#1}}}

\newcommand{\reftab}[1]{{Table~\ref{#1}}}

\hypersetup{hyperindex,breaklinks,colorlinks=true,linkcolor=blue,citecolor=blue,urlcolor=blue,filecolor=blue}

\begin{document}

\title{Effective Hamiltonians for the study of real metals using quantum chemical theories}

\author{Tina~N.~Mihm$^{(a),(b)}$}
\author{Tobias~Sch\"afer$^{(c)}$}
\author{Sai~Kumar~Ramadugu$^{(a),(b)}$}
\author{Andreas~Gr\"uneis$^{(c)}$}
\author{James~J.~Shepherd$^{(a),(b)}$}
\email{james-shepherd@uiowa.edu}
\address{$^{(a)}$ Department of Chemistry, University of Iowa, Iowa City, Iowa, 52242, USA \\ $^{(b)}$ University of Iowa Informatics Initiative, University of Iowa, Iowa City, Iowa, 52242, USA \\ $^{(c)}$ Institute for Theoretical Physics, TU Wien, Wiedner Hauptstraße 8-10/136, 1040 Vienna, Austria}

\begin{abstract}

Computationally efficient and accurate quantum mechanical approximations to solve the many-electron Schr\"odinger
equation are at the heart of computational materials science.
In that respect the coupled cluster hierarchy of methods plays a central role in molecular quantum chemistry
because of its systematic improvability and computational efficiency. 
In this hierarchy, coupled cluster singles and doubles (CCSD) is one of the most important
steps in moving towards chemical accuracy and, in recent years, its scope has successfully been expanded to the study of insulating surfaces and solids.
Here, we show that CCSD theory can also be applied to real metals. 
In so doing, we overcome the limitation of needing extremely large supercells to capture long range electronic correlation effects.
An effective Hamiltonian can be found using the transition structure factor 
-- a map of electronic excitations from the Hartree--Fock wavefunction -- which has fewer finite size effects
than conventional periodic boundary conditions.
This not only paves the way of applying coupled cluster methods to real metals but also reduces the computational
cost by two orders of magnitude compared to previous methods.
Our applications to phases of lithium and silicon show a resounding success in reaching the thermodynamic
limit, taking the first step towards a truly universal quantum chemical treatment of solids.
\end{abstract}
\date{\today}
\maketitle

\begin{figure*}
\subfigure[\mbox{}]{%
\includegraphics[width=0.45\textwidth,height=\textheight,keepaspectratio]{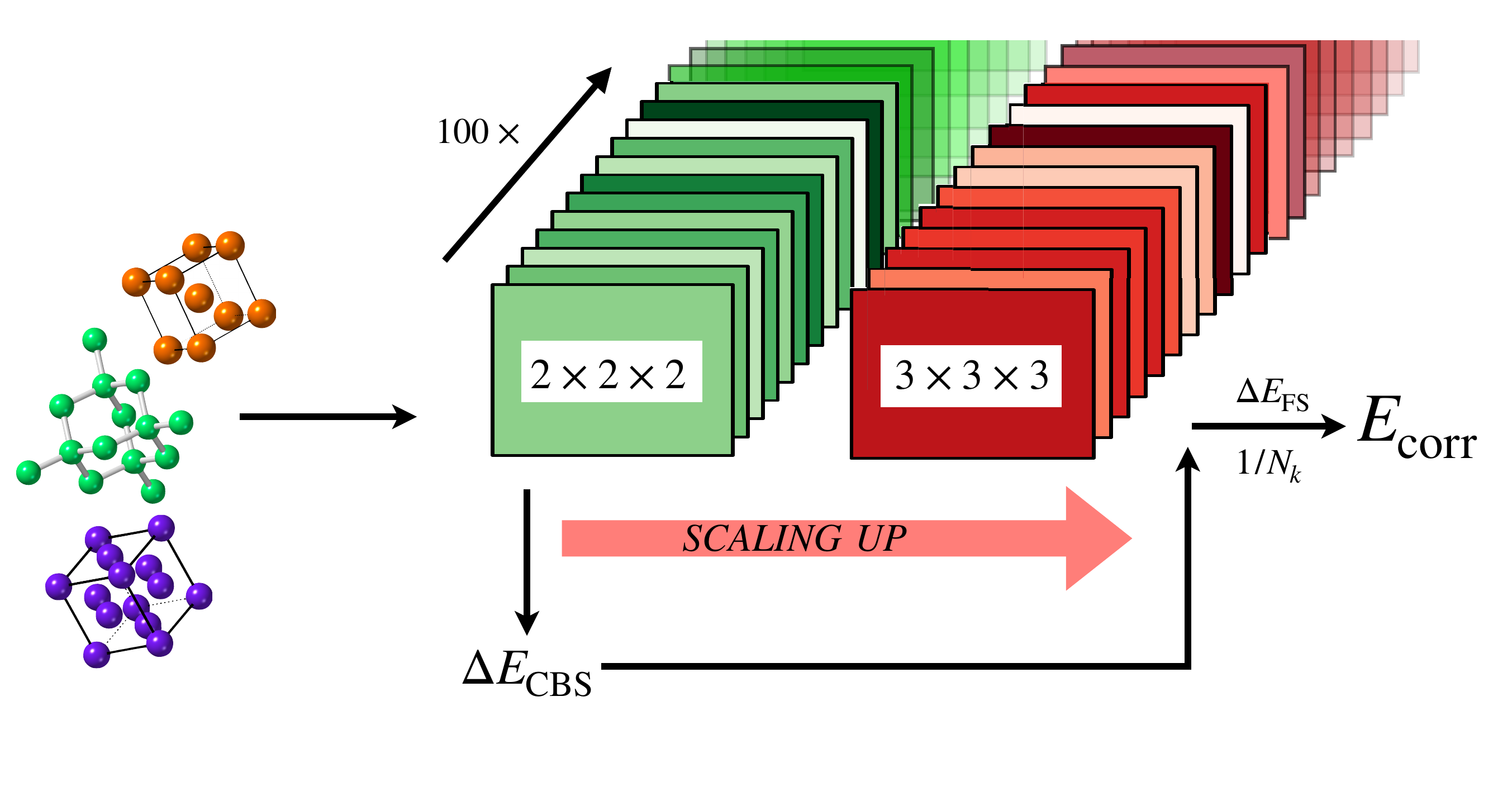}
\label{subfig:TA_flow}
}\quad\quad
\subfigure[\mbox{}]{%
\includegraphics[width=0.45\textwidth,height=\textheight,keepaspectratio]{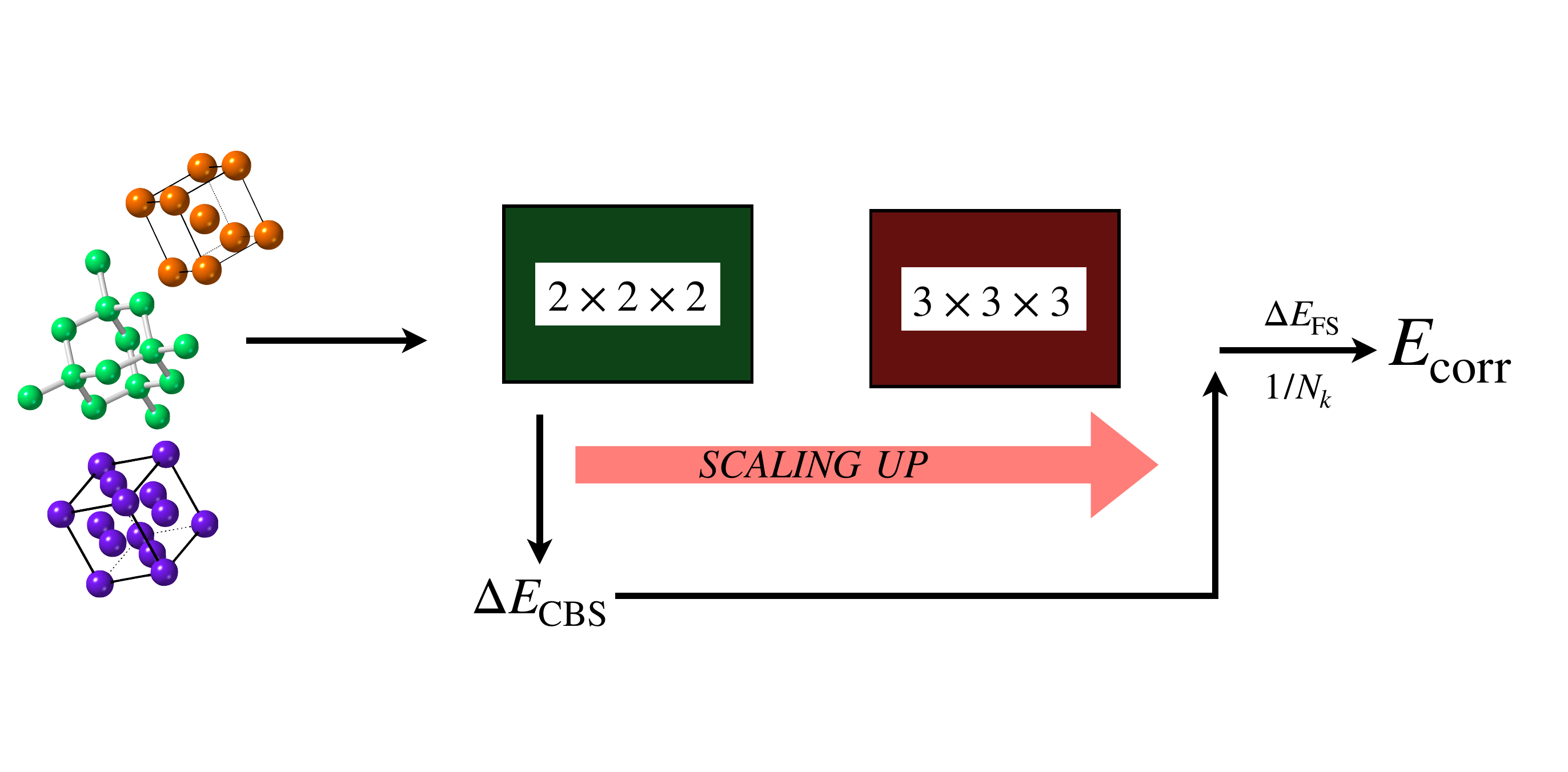}
\label{subfig:sfTA_flow}
}
\subfigure[\mbox{}]{%
\includegraphics[width=0.45\textwidth,height=\textheight,keepaspectratio]{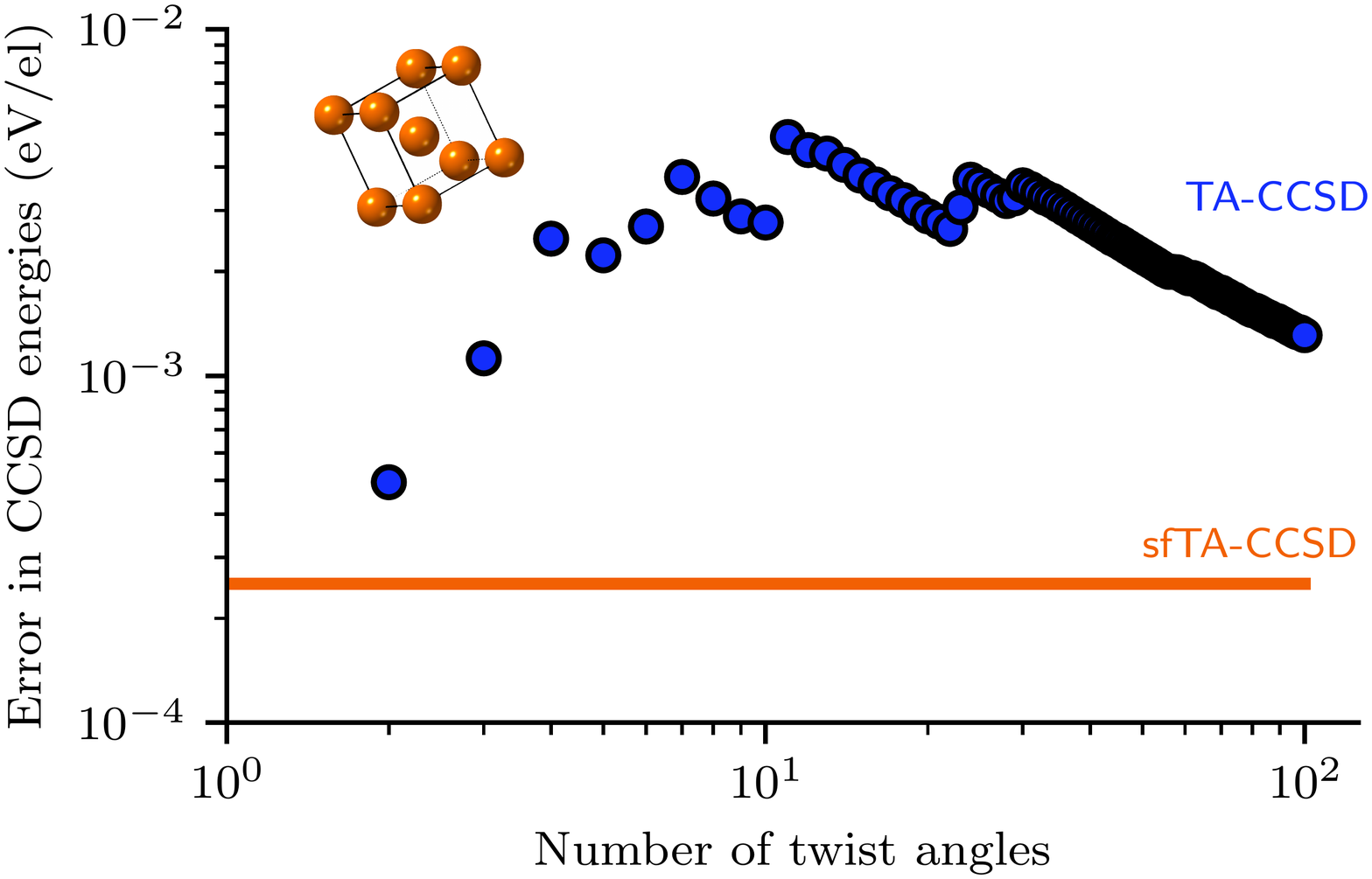}
\label{subfig:Error_in_TA}
}
\caption{Diagram to show algorithmic details and computational efficiency of the method proposed here, structure factor twist averaging (sfTA), in comparison to the conventional method, twist averaging. 
For the twist averaging method in (a), $100$ CCSD calculations are run at random twist angles over a set grid size for a chosen solid. From these, the TA-CCSD energies are calculated and corrections are applied to obtain the final CCSD correlation energy.
For the sfTA method in (b), a single twist angle is selected based on how close to the average structure factor of the system it is. This twist angle is then used to run a single calculation on the chosen solid and grid size before the same corrections are applied to obtain the final correlation energy. In both methods, a basis set correction ($\Delta E_\mathrm{CBS}$) is calculated from the smaller grid size and applied to all grids. To correct finite size error either a correction is calculated ($\Delta E_\mathrm{FS}$), or an extrapolation to the TDL can be performed.         
In (c), the stochastic error in the TA-CCSD energy for a $32$ atom supercell of Na is shown for an increasing number of twist angles.
This error is not converged (indicating undersampling) for fewer than $40$ twist angles.  Here, the TA-CCSD energy at $100$ twist angles was used as the reference value to calculate these errors.
When our sfTA scheme is used, the systematic error is of the order $1$ meV/el. By contrast, the stochastic error in twist averaging only reached this value when over $100$ twist angles have been used, noting here that the reference value used to obtain these error is the TA-CCSD energy at $100$ twist angles. 
This comparison is evidence for a $100$-fold increase in computational efficiency in the CCSD calculation. 
 }
\label{fig:TwstAngleError}
\end{figure*}

Many-electron perturbation theories are at the heart of theoretical quantum chemistry and play a central role
in understanding and predicting chemical processes in the gas phase, condensed matter systems, and on surfaces.
Although finite-order perturbation theories were proposed initially~\cite{moller_note_1934}, %
it has become evident that renormalized perturbation theories---resummations of certain diagrams in the many-electron perturbation
series expansion---often exhibit a better trade-off between accuracy and computational cost, and are numerically more stable~\cite{coester_bound_1958, coester_time_1958, krotscheck_coupled-cluster_1980, kummel_many-fermion_1978, cizek_coupled_1980}.
The resulting theories have substantially contributed to the success of modern \emph{ab initio} methods. In particular, these include the coupled cluster
method and its different approximations, such as the widely-used coupled cluster singles and doubles plus perturbative triples theory (CCSD(T))~\cite{raghavachari_fifth-order_1989}. %
Yet despite the great advances of modern coupled cluster theory approximations and their applications to molecules, surfaces, and solids %
it remains difficult to treat metallic systems at the same level of theory, with the exception of applications to simple model Hamiltonians such as
the uniform electron gas~\cite{ceperley_ground_1980, mcclain_spectral_2016, irmler_duality_2019, mihm_advanced_2020, mihm_optimized_2019, neufeld_study_2017, shepherd_coupled_2014, shepherd_communication:_2016, shepherd_many-body_2013, shepherd_range-separated_2014, spencer_developments_2016}. %
This constitutes a significant obstacle in modern computational materials science, where accurate benchmark results obtained with high-level theories
for all types of materials are desperately needed to complement computationally more-efficient but less-accurate density functional theory calculations. 
The important role of metals in surface chemistry, catalysis, and many technological applications highlights the necessity to expand the scope
of accurate quantum chemical perturbation theories further.
Therefore we are confident that the techniques outlined in this work, which make it possible to study metals using coupled cluster theory routinely, present a significant advance especially in combination with other methodological improvements such as embedding theories~\cite{lau_regional_2021, schafer_local_2021, chulhai_projection-based_2018}.

Despite recent successes of methods development~\cite{ booth_plane_2016, booth_towards_2013, gruber_applying_2018, white_time-dependent_2018, lewis_ab_2019, liao_communication:_2016, wang_excitons_2020, gillan_high-precision_2008,schafer_quartic_2017}, %
metals have still remained a formidable challenge to quantum chemistry.
The quintessential property of a metal---that it has a zero gap---has led to the view of these calculations being intractable.
This view comes from experiences in molecular systems with small HOMO-LUMO gaps, which are prone to strong correlation effects that are inaccessible by single reference methods like coupled cluster theories.
Further, small gaps can lead to numerical instability in both Hartree--Fock and resummation methods just in the regime where one approaches the thermodynamic limit~\cite{mcclain_gaussian-based_2017,gruber_applying_2018, gruneis_second-order_2010, lowy_electron_1975}. %
However, for the uniform electron gas, which is the simplest and most popular
metallic model Hamiltonian, each of these barriers can be overcome. \cite{mcclain_spectral_2016, neufeld_study_2017, shepherd_coupled_2014, shepherd_communication:_2016, shepherd_many-body_2013, shepherd_range-separated_2014, spencer_developments_2016}
Through scrupulous accounting for error and its cancellation, we have found that it is possible to converge each component of the total energy to reach benchmark accuracy.
On the journey to a universal application of single-reference coupled cluster theory to all solids, the final hurdle is to transfer these successes to real metals. 
Here, we present a simple but efficient scheme to determine an effective Hamiltonian that directly enables the application
of widely-used coupled cluster methods to study real metals. 
The proposed method determines a crystal momentum vector used in the periodic boundary
conditions of the \emph{ab initio} calculations, such that the electronic transition structure factor in the thermodynamic limit
is best approximated by a single finite periodic system. Our method ensures that degeneracies in the ground
state of the unperturbed Hartree--Fock Hamiltonian are automatically lifted, which is crucial for the numerical stability of these
self-consistent field calculations of metals. The resulting Hartree--Fock ground state is used as a reference for the coupled cluster theory
calculation in a numerically-stable and efficient manner. We apply the outlined method to prototypical metallic lithium and silicon phases.

\begin{figure}
\includegraphics[width=0.45\textwidth,height=\textheight,keepaspectratio]{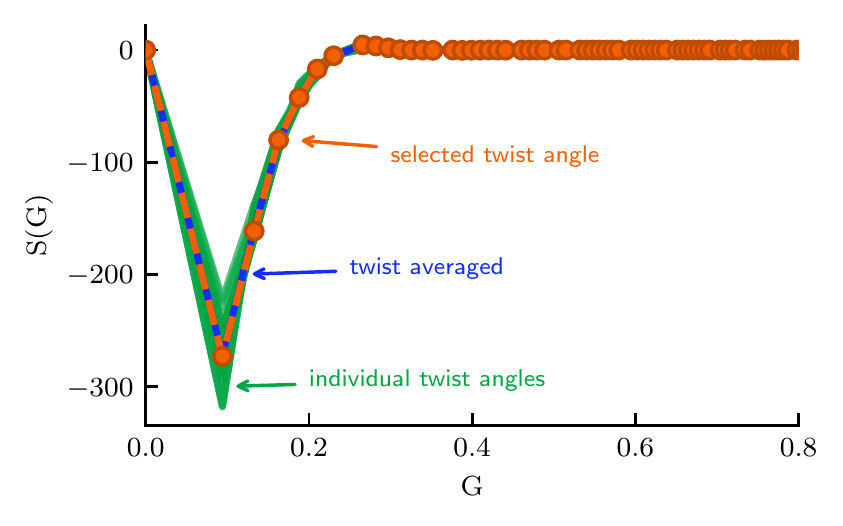}
\label{subfig:TA+cTA_CCSD}
\caption{A comparison between the full twist averaged transition structure factor and the sfTA structure factor as computed from the special twist angle for a $32$ atom Na system using CCSD. 
As can be seen in the graph, the sfTA structure factor appears to lie directly on top of the twist averaged structure factor (MAE = $0.08(13)$)
This demonstrates the ability of the method to reproduce the TA-CCSD results for solids.  
}
\label{fig:SF-TA-CCSD_Na}
\end{figure}

\section{Results}

The quantum mechanical many-electron Hamiltonian of periodic solids is invariant under any translational symmetry transformation that respects the periodicity of the attractive nuclear potential.
Crystal momentum vectors ${\bf k}$ that lie in the first Brillouin zone serve
as a label for translational symmetry transformations.
The full Hamiltonian can be written as
\begin{equation}
\hat{{H}}= \int \mathrm{d}{\bf k} \left \{ \hat{h}_1({\bf k}) + \iiint \mathrm{d}{\bf k}' \mathrm{d}{\bf k}''\mathrm{d}{\bf k}''' \hat{h}_2\left({\bf k},{\bf k}',{\bf k}'',{\bf k}'''\right) \right \},
\label{eq:H}
\end{equation}
where the one- and two-electron Hamiltonians for a particular set of $k$-vectors
are given in second quantization by
\begin{equation}
\hat{h}_1({\bf k}) = \sum_{p({\bf k}), q({\bf k})}  h_{pq} a^\dagger_{p} a_{q}
\end{equation}
and
\begin{equation}
\hat{h}_2({\bf k},{\bf k}',{\bf k}'',{\bf k}''')
= \frac{1}{2} \sum_{p({\bf k}), q({\bf k}')} \sum_{r({\bf k}''), s({\bf k}''') }
 \langle pq | rs \rangle a^\dagger_{p} a^\dagger_{q} a_{r} a_{s},
\end{equation}
respectively. The indices $p,q,r$ and $s$ refer to occupied and unoccupied one-electron
Bloch states characterized by crystal momentum vectors $\bf k$. The two-electron integrals $\langle pq | rs \rangle$
are non-zero only if the corresponding $k$-vectors conserve total crystal momentum.
In practice, the thermodynamic limit (TDL)%
, computed by integrating over all $\bf{k}$, 
is approached by sampling the
first Brillouin zone using a regular $k$-mesh with an increasing number of $k$-points
($\lim_{N_k \to \infty}  \sum_{i=1}^{N_k} \frac{1}{N_k \Omega}$ ).

Here, we seek to define an effective Hamiltonian that can also be applied to metals and, in combination with recently proposed finite size corrections~\cite{gruber_applying_2018, gruber_ab_2018}, %
yields an even more rapid and numerically-stable convergence of correlation energies to the TDL.
To this end, we approximate the Brillouin zone integrations in \refeq{eq:H} such that
$\int d{\bf k} F({\bf k}) \approx  \sum_{i=1}^{N_k} \frac{1}{N_k \Omega} F({{\bf k}}_i+{{\bf k}_s})$. Here, F({\bf k}) refers to the terms inside the curly brackets in \refeq{eq:H}.
The vector ${\bf k}_s$ corresponds to an offset of the employed $k$-mesh with respect to the origin,
yielding a coupled cluster correlation energy dispersion.
Current approaches take an average over different offsets,
which increases the computational cost of a calculation by a significant factor.
Here, we find that an effective Hamiltonian can be formed and the simulation run at one specific offset only, which we call the effective twist angle. 

Our approach is to select one effective twist angle, or ${\bf k}_s$, at which to
perform the calculation by considering the twist angle that best represents the electronic transition
structure factor within an first-order M{\o}ller-Plesset (MP1) wavefunction formalism. %
This is represented schematically in \reffig{fig:TwstAngleError}. 
The Fourier components of the transition structure factor can be used to express
the projected correlation energy obtained from a wavefunction $| \Psi \rangle$
such that
\begin{equation}
E_\mathrm{corr}=\langle \Psi_0 | H-E_0 | \Psi \rangle = {\sum_{{\bf G}}}' {v}({\bf G}) {S}({\bf G}).
\label{eq:ecorrg}
\end{equation}
Here, the momentum ${\bf G}$ corresponds to a plane wave vector that is
defined as ${\bf G}={\bf g} + {\Delta \bf k}$, where ${\bf g}$ is a reciprocal lattice vector
and ${\Delta \bf k}$ is the difference between any two crystal momentum vectors that are conventionally
chosen to sample the first Brillouin zone.
$v({\bf G})$ are the Fourier coefficients of Coulomb kernel with the familiar form of $\frac{4\pi}{G^2}$ for
excitations allowed by momentum conservation and the prime on the sum implies that the G=0 contribution is treated in an approximate fashion.\cite{liao_communication:_2016}

Our approach is to determine an effective twist angle that most closely approximates the twist-averaged
structure factor ($\overline{S}({\bf G})=\sum_{{\bf k}_s} \frac{1}{N_{{\bf k}_s}}{S}_{{\bf k}_s}({\bf G})$)
from MP2 theory, {\it i.e.}, minimizes the residual 
$\sum_{\bf G} \left |  \overline{S}({\bf G}) - S_{{\bf k}_s}({\bf G}) \right |^2$.
This approach is based on the idea that the single twist angle then represents the average structure factor, not just
for MP2, but for other methods as well. 
The single twist angle is then used to calculate the correlation energy using CCSD. 
An example of this is illustrated in \reffig{fig:SF-TA-CCSD_Na} and the two structure factors are visually
indistinguishable from one another. %
This small error demonstrates that one twist angle is able to reproduce the transition structure factor averaged across ${\bf k}_s$.
As this method is a way to produce the twist averaged energy based on the best representation of the structure factor, we refer to it as \emph{structure-factor-based twist averaging} (sfTA). 

Our protocol to determine the total energy and energy differences between material phases is: (a) run sfTA CCSD calculations at increasingly dense $k$-point meshes keeping the number of orbitals per $k$-point constant; (b) for a small $k$-point mesh (e.g. $2 \times 2 \times 2$) converge to the complete basis set limit using natural orbitals and apply the following correction: $\Delta E_{\mathrm{CBS}} (222) = E_{\mathrm{CBS}} (222)  –  E_{\mathrm{corr}}(M,222)$; \emph{either} (c) apply finite size corrections to each mesh using an interpolation scheme~\cite{gruber_applying_2018}; \emph{or} (d) extrapolate the correlation energy to the TDL using a power-law of $1/N_k$; (e) to find the total energy, the correlation energy from steps (a)-(d) is added to Hartree--Fock calculations which have been extrapolated to the TDL. As HF is significantly less expensive than CCSD, larger $k$-point meshes (up to $6 \times 6 \times 6$) are used for this step.

\begin{figure}
\includegraphics[width=0.45\textwidth,height=\textheight,keepaspectratio]{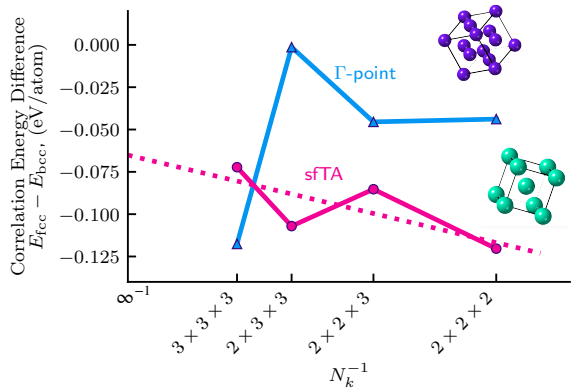}
\caption{
Comparison between $\Gamma$-point and sfTA correlation energy for the two phases of lithium. 
The comparison, shown as a difference between the face-centered cubic (fcc) and body-centered cubic (bcc) phases, is graphed against k-points ranging from $2\times2\times2$ to $3\times3\times3$. All k-pionts have a basis set correction applied. 
As can be seen from this comparison, the sfTA results show a much smoother convergence of the difference with a difference of $-0.07(2)$ eV/atom at the TDL. 
The numbers in parentheses are the errors which come from the fitting algorithm.
}
\label{fig:Li_phase_diagram}
\end{figure}

\begin{table*}[]
{%
\caption{
Equilibrium lattice properties for the two silicon phases, including the transition volume ($V_t$), the Bulk modulus ($B_0$), pressure derivative of the Bulk modulus ($B'_0$), and the volume at equilibrium ($V_0$). The energy difference was calculated as the difference between the minimum of the curves for $\beta$-Sn and diamond structures. 
These fits were used to find the transition pressure ($P_t$).  
Our HF and CCSD are shown in comparison to previous quantum Monte Carlo studies~\cite{hennig_phase_2010, purwanto_pressure-induced_2009, alfe_diamond_2004, maezono_diamond_2010}.
We have corrected all of the energy differences except HF and DMC+EMP-pp to include a core-polarization correction of $30$ meV/atom taken from Ref.~\onlinecite{alfe_diamond_2004}; these are also applied to the transition pressures. 
For DMC+EMP-pp, we use the data including the core-polarization corrections referenced in their paper. Also, the DMC+EMP-pp contains a correction to the energy-volume curve using the empirical (EMP) pseudopotential. 
Finally, for all methods, the fully corrected transition pressures contain the corrections for zero point energy and finite-temperature vibrational effects and core-polarization corrections. 
We include these vibrational correction terms in CCSD following the numbers from Ref.~\onlinecite{alfe_diamond_2004}.
Experimental numbers were given by Ref.~\onlinecite{hull_properties_1999} and Ref.~\onlinecite{mcmahon_pressure_1994} for diamond and $\beta$-Sn respectively as referenced in Ref.~\onlinecite{hennig_phase_2010}.
The experimental range for the transition pressure comes from Ref.~\onlinecite{hu_crystal_1986}.
}\label{table1}
\vspace{10pt} \begin{tabular}{llccccccc}
Structure     & Property                           & HF    & CCSD-FS & Expt.       & DMC \footnote{\label{fn:a} Alfe \emph{et al.} 2004 ~\cite{alfe_diamond_2004}} & DMC \footnote{\label{fn:b} Hennig \emph{et al.} 2010~\cite{hennig_phase_2010}} & DMC+EMP-pp \footnote{\label{fn:d} Maezono \emph{et al.} 2010~\cite{maezono_diamond_2010} } & AFQMC                                                              \\
\hline
\hline
D-Si          & $V_t$ ($\AA^3$/atom)               & 15.51 & 17.29   & 18.15       & --    & 18.14 & 17.83      & 18.15                                                              \\
              & $B_0$ (GPa)                        & 102.5 & 101.1     & 99.2        & 103.0 & 98.0  & 96.2       & --                                                                 \\
              & $B'_0$                             & 3.88  & 4.0     & 4.11        & --    & 4.6   & 4.19       & --                                                                 \\
              & $V_0$ ($\AA^3$/atom)               & 20.79 & 20.15   & 20.0        & 20.11 & 19.98 & 19.75      & --                                                                 \\
$\beta$-Sn Si & $V_t$ ($\AA^3$/atom)               & 12.14 & 13.5    & 13.96       & --    & 13.9  & 13.81      & 13.955                                                             \\
              & $B_0$ (GPa)                        & 109   & 125.2    & --          & 114   & 107   & 104.2      & --                                                                 \\
              & $B'_0$                             & 4.17  & 3.9     & --          & --    & 4.6   & 4.7        & --                                                                 \\
              & $V_0$ ($\AA^3$/atom)               & 15.97 & 15.20   & --          & 15.26 & 15.2  & 15.17      & --                                                                 \\
\hline
\hline
              & $\Delta$ E (eV/atom)               & 1.303 & 0.609   & --          & 0.505 & 0.424 & 0.329      & 0.365 \footnote{\label{fn:d} core-polarization correction applied by us} \\
              & $P_t$ (GPa)                        & 52.98 &  20.94   & --          & 17.8  & 15.3  & 13.16      & 13.9                                                               \\
              & $P_t$ incl. vib. corrections (GPa) & 51.68 & 19.64   & 11.3 - 12.5 & 16.5  & 14.0  & 12.2       & 12.6  \\
\hline
\end{tabular}%
}
\end{table*}

Figure \ref{fig:Li_phase_diagram} shows the difference between the correlation energies for the face-centered cubic (fcc) and body-centered cubic (bcc) phases of lithium. 
Calculations were performed on a variety of k-point meshes for extrapolation to the TDL. The corrected energies were then extrapolated to the TDL using a power law of $1/N_k$, where $N_k$ is the total number of $k$-points.
For both phases, a correction for the basis set incompleteness error was calculated based on the complete basis set (CBS) limit extrapolation for a $2\times2\times2$ k-point mesh and a range of basis sets as described in the protocol. 

After extrapolation to the TDL shown in \reffig{fig:Li_phase_diagram} was performed, the Hartree--Fock energy obtained from extrapolating TA-HF energies to the TDL (using k-point meshes up to $6\times6\times6$) was added.
Overall, this resulted in a total energy difference of $-0.06(5)$ eV/atom between fcc and bcc. 
This number indicates that the electronic contribution at $0$K does not change the relative stability of the bcc versus fcc phases of lithium within the remaining extrapolation error. 
Overall, the smoothness of this convergence to the TDL seen in \reffig{fig:Li_phase_diagram} when compared with the $\Gamma$-point calculation shows a successful application of sfTA.

\begin{figure}
\includegraphics[width=0.45\textwidth,height=\textheight,keepaspectratio]{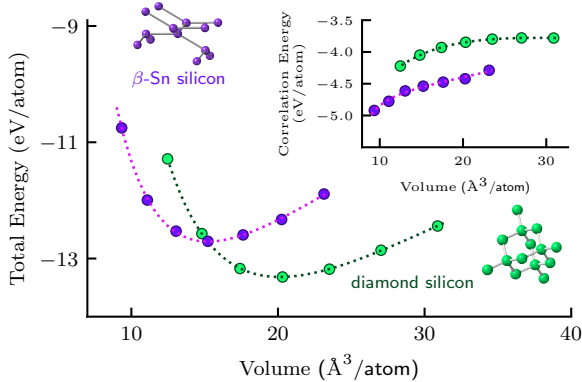}
\caption{Total energies for the diamond and $\beta$-tin phases of Si across a range of volumes. 
We can clearly see the minima for both phases, with diamond reaching a minimum at $20.0(4) \mathrm{\AA}^3$/atom and $\beta$-tin reaching a minimum at $15.97(7) \mathrm{\AA}^3$/atom. 
The transition between the two phases can clearly be seen to take place between the volumes around $15$ to $18 \mathrm{\AA}$/atom.
The inset on the graph shows the correlation for both phases for the same range of volumes.
The calculations shown were initially run using a $3\times3\times3$ k-point mesh. Then, finite size and basis set corrections were applied. 
 The $\beta$-Sn energies have a $30$~meV/atom correction added from the core polarization potential (CPP).
 These represent the best quality estimates we have for the complete basis set/thermodynamic limit energies for CCSD that we can compute.
}
\label{fig:Si_Vol_diagram}
\end{figure}

We now apply this method to the volume-energy curves of silicon in a diamond and $\beta$-tin lattice.
These allotropes of silicon are semiconducting and metallic, constituting an excellent test case for the ability of the proposed sfTA technique to treat small-gap systems on the same footing. 
Furthermore these phases have been studied extensively also by other electronic structure theories including DFT and the random phase approximation (RPA)~\cite{xiao_structural_2012}.

We performed calculations using sfTA-CCSD on a $3 \times 3 \times 3$ $k$-point mesh.
Finite size corrections were applied using a mesh-interpolation scheme and a complete basis set correction was derived from a $2 \times 2 \times 2$ $k$-point mesh run at the equilibrium volume.~\cite{gruber_applying_2018}
The sfTA-CCSD correlation energies were then added to TDL extrapolated TA-HF energies (using $k$-point meshes up to $5\times5\times5$) to obtain total energies.

The Birch-Murnaghan equation of state was fit to the diamond and $\beta$-Sn phases of silicon, shown in \reffig{fig:Si_Vol_diagram}, allowing us to obtain equilibrium lattice properties, including equilibrium volumes and bulk moduli. 
We find that the sfTA-CCSD energy curves are smooth both for the total and correlation energies. 
The fitting parameters are shown in \reftab{table1} compared to several high-accuracy quantum Monte Carlo methods
from previous studies~\cite{hennig_phase_2010, alfe_diamond_2004, purwanto_pressure-induced_2009, maezono_diamond_2010}.
Experimental values are also included~\cite{hu_crystal_1986,hull_properties_1999,mcmahon_pressure_1994}.

We note that the DMC findings compiled in \reftab{table1} span a relatively large range for different equilibrium properties including equilibrium volumes, bulk moduli, and its derivative. In particular the equilibrium volumes of
silicon diamond range from $19.75$~$\mathrm{\AA}^3$/atom to $20.11$~$\mathrm{\AA}^3$/atom.
The experimental equilibrium volume was measured to be $20.0$~$\mathrm{\AA}^3$/atom.
The difference in the DMC estimates can partly be explained by different
approximations used to correct finite size errors and the dependence on the employed pseudopotentials.
Similar trends as seen for the equilibrium volumes can also be observed for the bulk
moduli.
The most striking changes between the different DMC calculations can be observed
for the energy differences between both phases, which reduces from $505$~meV/atom in 
Ref.~\onlinecite{alfe_diamond_2004} to $329$~meV/atom in Ref.~\onlinecite{maezono_diamond_2010}.
Due to the strong dependence of the transition pressure between the diamond to 
$\beta$-Sn allotropes on the energy difference, the predictions change from
$17.8$~GPa to $13.16$~GPa, which is gradually improving compared to the experimental
estimates that lie between $11.2$ and $12.6$~GPa.
Despite the evolutionary character of the DMC findings summarized in \reftab{table1}, we consider the latest DMC calculations from Ref.~\onlinecite{maezono_diamond_2010} to be the literature benchmark that can be used
to assess other electronic structure calculations directly without the necessity to account for
other effects that need to be considered when comparing to experiment, such as
finite temperature and vibrational entropy contributions.
This is corroborated by an independent study using AFQMC which found numbers in good agreement with the spread of DMC results~\cite{purwanto_pressure-induced_2009}.

We now turn to the discussion of results obtained using the quantum chemical methods.
In this regard, HF forms the starting point and will also serve as reference for the
subsequent CCSD calculations.
The HF results listed in \reftab{table1}
demonstrate that HF theory strongly overestimates the diamond equilibrium volume and bulk
modulus compared to experiment and DMC, which can mainly be attributed to the neglect of electronic correlation effect. Likewise, the energy difference between the diamond
and $\beta$-Sn phase is significantly overestimated, yielding a transition pressure
that is almost five times larger than the experimental findings.

CCSD theory is able to substantially improve upon the HF description by accounting for
correlation effects once we have accounted for finite size as well as basis set corrections.
While the CCSD equilibrium volume and bulk modulus for the diamond phase are in
good agreement with experiment and latest DMC estimates, we find that the
predicted transition pressure are still overestimated by several GPa.
Our calculated CCSD energy difference between the two phases was found to be $0.609$~eV/atom, with a transition pressure of $20.94$ GPa
as seen in \reftab{table1}.
This energy difference was found to be larger than the largest DMC energy difference ($0.505$~eV/atom) by $0.104$ eV/atom.
This is promising because CCSD is the simplest correction to HF in the coupled cluster hierarchy of methods and these calculations took only $4$ days (per volume point) on $16$ cores. %

The remaining discrepancy between CCSD and DMC is not surprising.
Although CCSD forms an important step towards chemical accuracy, it is known
from quantum chemical calculations of molecules that the inclusion of perturbative triple
particle-hole excitation operators, as performed by CCSD(T), is needed for chemically accurate
reaction energies~\cite{helgaker_priori_2004}. 
However, we have previously shown that the method is divergent due to its perturbative component~\cite{shepherd_many-body_2013}.
One alternative possibility to improve upon CCSD theory has recently arisen in the literature with the distinguishable cluster theory (DCSD)~\cite{kats_communication_2013}.
When we run this at the mid-point volume we find that DCSD lowers the energy gap between the phases by $68$ meV/atom to $541$ meV/atom.
If this were uniform across the volume curve, this is equivalent to lowering the transition pressure to $18.6$ GPa.

Taken together, these numbers are in reasonable agreement with the DMC energies. 
The improvement from CCSD to DCSD also demonstrates the way in which quantum chemical methods can be improved. 
In closing, we do agree with the authors of the DMC study in Ref.~\onlinecite{alfe_diamond_2004}, who conclude that the numbers from electronic structure calculations were only in agreement with experiment after accounting for vibrational and anharmonic effects.
In practice, this means that the structural transition pressure should be a little higher than experiment if the electronic structure is being treated appropriately.

\section{Discussion}

In conclusion, we developed a new method to determine an effective Hamiltonian for
treating metallic systems using quantum chemical methods.
This method, which we call sfTA, finds the single twist angle which best reproduces the twist-averaged transition structure factor.
Then, a single calculation can be used to compute an energy that is very close to the twist-averaged CCSD energy with significant cost reduction.
The selected twist angle can be found efficiently regardless of the band gap of the system. 
Our proof-of-principle results come from studying the phases of lithium and silicon.
Here, we used the effective Hamiltonian in conjunction with several other recently-developed methods to converge the calculations with respect to basis set size and particle number. 
The fcc and bcc lithium phases were found to be degenerate with each other, indicating relative stabilities between the phases in the thermodynamic limit. 
For silicon, we found results that were in reasonable agreement with literature benchmarks from quantum Monte Carlo. %
We found that the distinguishable cluster approximation reduced the energy gap between the two phases resulting in agreement with DMC energies.
Overall, these results demonstrate that our method is capable of obtaining good CCSD results for a range of metallic applications.

Coupled cluster theory benefits from such an approach more than quantum Monte Carlo for two significant reasons.
The first is that basis set incompleteness error from a twist-averaged (or balanced) description of correlation commutes more reliably with the electron number. This provides the crucial benefit of reducing the basis set size required to treat metals and materials with small band gaps. %
The second is that TDL corrections depend upon the same relationship, and the increased commutivity also helps these corrections become more consistent. 
Taken together, the small change in the Hamiltonian's symmetry affects the calculation dramatically: calculations that previously were failure-prone are now routinely feasible.

\section{Methods}

{\bf\emph{Codes and methods:-}} All calculations were performed using {\tt{VASP}} \cite{kresse_norm-conserving_1994, kresse_efficient_1996} and the projector augmented-wave (PAW) method in a plane wave basis set~\cite{blochl_projector_1994}.
The corrections for the finite size effects for lithium and silicon were performed using cc4s interfaced with the {\tt{VASP}} code~\cite{gruber_applying_2018}. %
Twist averaging was used to obtain the Hartree--Fock data. 
All other calculations were run using the sfTA method.
Canonical Hartree--Fock (HF) orbitals were used for all MP2 calculations.
For coupled cluster, we used approximate natural orbitals,  estimated from MP2 natural orbitals~\cite{gruneis_natural_2011}, for the lithium basis set convergence calculations and all of the silicon phase calculations.%
The rest of the calculations used canonical HF orbitals.

{\bf\emph{Calculation details:-}} 
We used a plane wave energy cutoff ({\tt{ENCUT}}) of $400$~eV throughout. %
For the initial/preliminary sodium calculations (and the carbon and silicon calculations in the SI), we used an auxilliary plane wave basis set cutoff ({\tt{ENCUTGW}}) of $150$~eV. The basis set was truncated to $48$ orbitals per k-point (using the {\tt{NBANDS}} input); these calculations were all $1\times1\times1$ supercells. 
For lithium, these variables were used to perform calculations at k-point meshes of $2\times2\times2$, $2\times2\times3$, $2\times3\times3$, and $3\times3\times3$. 
Silicon followed likewise with the original calculations at $3\times3\times3$. 

In order to calculate the complete basis set energy for each lithium and silicon phase, {\tt{NBANDS}} was changed to give $16$, $24$, $32$, $40$ and $48$ orbitals per k-point. 
For these calculations, {\tt{ENCUTGW}} was set to $400$~eV in order to ensure the initial basis set was large enough that the number of orbitals designated by the {\tt{NBANDS}} input was the limiting factor for truncating the basis set in all calculations.
The lithium bcc calculations used a different {\tt{NBANDS}} of $24$ orbitals per k-point in the HF calculation preceding the approximate natural orbital calculation (when varying the {\tt{NBANDS}} used in the coupled cluster part of the calculation) which is the same number of k-points per atom as fcc. 

To obtain converged energies for the CCSD and MP2 calculations, an energy difference of $1\times10^{-4}$~eV was used for the Li, Na and C calculations. The Si phases required a smaller energy difference of $1\times10^{-6}$~eV  to obtain fully converged energies.

{\bf\emph{Structure factor:-}} 
We introduced the structure factor in the main manuscript in \refeq{eq:ecorrg}, where it was defined in relation to the coupled cluster wavefunction. 
In describing the structure factor below, we follow the notation of Ref.~\onlinecite{liao_communication:_2016} for mathematical consistency.%
The CC wavefunction has amplitudes which can be written ${\it T}_{ij}^{ab}={\it t}_{ij}^{ab}+{\it t}_{i}^{a}{\it t}_{j}^{b}$. Here, the $i$ and $j$ indices refer to occupied orbitals while $a$ and $b$ indices refer to unoccupied orbitals. %
Then ${\it t}_{i}^{a}$ and ${\it t}_{ij}^{ab}$ are singles and doubles amplitudes, respectively.%
Then the transition structure factor can be constructed as: 
\begin{equation}
\begin{split}
S({\bf G})=\sum_{{\bf k}_i,{\bf k}_j,{\bf k}_a}\sum_{n_i,n_j,n_a,n_b}{\it T}_{ij}^{ab} 
[ & 2C_i^a({\bf G})C_b^{j*}({\bf G}) \\ & -C_i^b({\bf G})C_a^{j*}({\bf G}) ],
\end{split}
\end{equation}
The coefficients $C_i^a({\bf G})$ arise from the Fourier transform of the co-densities of two orbitals $i$ and $a$ and the electron repulsion integrals can also be written in terms of these intermediates~\cite{liao_communication:_2016}.

{\bf\emph{Lattice constants:-}} The equilibrium lattice constants for $\beta$-tin silicon and diamond silicon (which are $4.9$~\AA~ and $5.761$~\AA~ respectively) were scaled by factors in the range $0.85$ to $1.1$ to produce a range of volumes. 
Lithium structure information, including equilibrium lattice constants, were obtained from the NOMAD Encyclopedia data base~\cite{scheffler_nomad_2015}.

{\bf\emph{Numerical analysis:-}} For the extrapolations to the TDL and the Birch-Murnaghan fits the {\tt{numpy}} and {\tt{scipy}} libraries were used with {\tt{Python 3.7.3}}. 
Error bars on the TDL extrapolations 
reflect standard error in the fit parameters. 

\section{Acknowledgements}

The University of Iowa is acknowledged for funding and computer time (JJS, SKR, TNM).
A.G. and T.S. thankfully acknowledge support and funding from the European
Research Council (ERC) under the European Unions
Horizon 2020 research and innovation program (Grant Agreement No 715594).
We acknowledge Michael Mavros for his verbal comments on the manuscript. 

\section{Contributions}

The intellectual origination of the presented sfTA method was at the University of Iowa (JJS, TNM). 
JJS and TNM conceived of, developed, and implemented the sfTA algorithm (i.e. wrote the code which performs twist angle selection).
TNM, TS, AG, and JJS contributed to the choice of systems to study, analyses, and interpretations.
AG leads the development of the coupled cluster routines used for the performed ab initio calculations.
TS contributed to the development and implementation of correlated routines used for performed ab initio calculations.
SKR installed, managed, and tested the software used.
All authors contributed to the running of calculations.
TNM, AG, and JJS wrote the manuscript.

\bibliography{tm_zotero}

 \end{document}